\documentclass[%
 reprint,
superscriptaddress,
 amsmath,amssymb,
 aps,
]{revtex4-1}

\usepackage{subcaption}

\usepackage{bm} 
\usepackage{graphicx}
\usepackage{dcolumn}
\usepackage{bm}
\usepackage{hyperref}

\begin{document}

\preprint{APS/123-QED}

\title{The Weak Cosmic Censorship Conjecture in Hairy Kerr Black Holes}

\author{Lai Zhao}

\author{Meirong Tang}

\author{Zhaoyi Xu}%
\email{zyxu@gzu.edu.cn(Corresponding author)}
\affiliation{%
 College of Physics,Guizhou University,Guiyang,550025,China
}%


\begin{abstract}
The Weak Cosmic Censorship Conjecture, since its proposal, has always been a controversial hypothesis, but its significance in astrophysics is undeniable. For a regular black hole, its center does not contain a singularity, and the destruction of the horizon of such black holes is not protected by the Weak Cosmic Censorship Conjecture. Therefore, we employ Gedanken experiments to study the hairy Kerr black holes, which are promising candidates to serve as "simulators" of astrophysical black holes. By investigating these black holes through testing particles and scalar fields carrying large angular momentum, we explore whether these black holes can achieve overspinning. Our results suggest that the overspinning behavior of these hairy Kerr black holes in   extremal or near-extremal conditions strongly depends on the hairy parameters (${\alpha, l_0}$). This not only potentially offers us an opportunity to explore the interior structure of black holes, but may also provide clues for constraining the hairy parameters. This phenomenon might reveal the connection between the no-hair theorem of black holes and the weak cosmic censorship conjecture, bringing new perspectives to our understanding of these theories.
\begin{description}
\item[Keywords]
Weak Cosmic Censorship Conjecture; Hairy Kerr Black Hole; Scalar Field; No-Hair Theorem for Black Holes.
\end{description}
\end{abstract}

\maketitle


\section{\label{sec:level1}Introduction}
Black holes, as a product of the predictions of general relativity, were most directly confirmed only recently when LIGO detected gravitational wave signals from a binary black hole merger for the first time\cite{LIGOScientific:2016emj}. Within the framework of general relativity, the singularity theorems proposed by Penrose and Hawking state that, under certain material energy and initial conditions, gravitational collapse inevitably leads to the emergence of spacetime singularities\cite{Penrose:1964wq,Hawking}. At these singularities, all physical laws diverge, meaning gravitational theory is not applicable near these singularities. If these exposed "points" (naked singularities) exist in spacetime, they would disrupt the well-defined spacetime geometry and causal laws. To ensure that physical laws are not threatened by these naked singularities, Penrose proposed a hypothesis that these so-called singular points should not exist independently in spacetime but should be surrounded by an event horizon. Only if these singular points are hidden within the horizon, the geometry of spacetime and causal laws outside the event horizon will not be affected by the singularity. This is what is known as the Weak Cosmic Censorship Conjecture (WCCC)\cite{Penrose}. The Weak Cosmic Censorship Conjecture, as a mechanism to protect classical gravitational theory, although it cannot be described by a strict mathematical formula to date\cite{Wald:1997wa}, its status in black hole physics is undeniably significant.

The Weak Cosmic Censorship Conjecture, serving as a hypothesis for classical gravitational theory, requires further exploration to determine its universal applicability. If this conjecture is violated, we may observe the internal structure of black holes and quantum gravitational phenomena\cite{Li:2013sea}. There are many methods to test the Weak Cosmic Censorship Conjecture, such as through numerical simulations, like simulating the collapse evolution of matter\cite{Christodoulou:1984mz,Joshi:2011rlc,Mizuno:2016onf,Ames:2023akp,Corelli:2021ikv,Tavakoli:2020xgc,Mosani:2020mro,Manna:2019tql,Sharif:2018mwx,Figueras:2015hkb}, collisions of supermassive black holes\cite{Andrade:2019edf,Andrade:2018yqu,Brill:1993tm,Choptuik:2003as}, and testing the conjecture in phase space\cite{Li:2020nnz,Hu:2020lkg,Han:2019lfs,Han:2019kjr,Zeng:2019huf}. Besides these methods, in 1974, Wald  first proposed a version of a thought experiment (Gedanken experiment)\cite{Wald:1974hkz}, considering throwing a test particle with large electric charge and angular momentum into an extremal Kerr-Newman (KN) black hole. The results showed that the Weak Cosmic Censorship Conjecture is supported in Kerr-Newman black holes under first-order perturbations. Since the first version of the thought experiment was proposed, corresponding situations have been discussed in different types of black holes\cite{Jacobson:2009kt,Saa:2011wq,Ghosh:2019dzq}. Notably, in reference\cite{Saa:2011wq}, Alberto Saa and others considered  near-extremal Kerr-Newman black holes under first-order perturbations. Their results indicated that a test particle could fall into the black hole, potentially creating a naked singularity by disrupting the event horizon. These studies focused on first-order perturbations. Later, Hubeny extended the first version of the thought experiment to second-order perturbations, showing that the event horizon of a Kerr-Newman black hole could be disrupted in extremal cases\cite{Hubeny:1998ga}. This led to a surge of research\cite{Gao:2012ca,Gwak:2015fsa,Siahaan:2015ljs,Shaymatov:2022ako,Ying:2020bch,Chirco:2010rq,deFelice:2001wj,He:2019kws,Chen:2019pdj,Matsas:2007bj,Richartz:2008xm,Rocha:2014jma,Duztas:2016xfg,Gwak:2017kkt}. Some of these studies suggest the possibility of black hole overspinning. Notably, in reference\cite{Gao:2012ca}, Gao and others mentioned that if spacetime effects are fully considered during the process of a particle falling into a black hole, the Weak Cosmic Censorship Conjecture seems to be supported. To make the Gedanken experiment more rigorous, in 2017, Sorce and Wald proposed a more complex second version\cite{Sorce:2017dst,Wald:2018xxi}, extending Hubeny's Gedanken experiment results to any matter satisfying zero-energy conditions\cite{Hubeny:1998ga,Iyer1}. Their results suggest that disrupting the event horizon of near-extremal Kerr-Newman black holes becomes improbable after considering certain factors. Following the new version of the Gedanken experiment, research in different systems or backgrounds emerged, such as\cite{Duztas:2019ick,Chen:2019nsr,Gwak:2019asi,Jiang:2019vww,Chen:2018yah,Gwak:2018akg,Figueras:2017zwa,Yang:2022yvq,Wang:2021kcq,Qu:2021hxh,Zhang:2020txy,Zhao:2023vxq,Meng:2023vkg,Feng:2020tyc}, where authors tested the Weak Cosmic Censorship Conjecture in various scenarios.

Recently, explorations into quantum effects have sparked significant interest in Gedanken experiments. In two versions of the Gedanken experiments, considering spacetime effects and radiation effects seems to solidify the status of the Weak Cosmic Censorship Conjecture. However, in an article by Li Zilong et al, they suggest that, under quantum effects, it is possible for the event horizon of a regular black hole to be disrupted\cite{Li:2013sea}. In other words, regular black holes might not be under the "supervision" of the Weak Cosmic Censorship Conjecture, and their event horizons could be disrupted, opening possibilities for studying the internal structure of black holes. Considering some quantum tunneling scenarios, the disruption of the event horizon is possible\cite{Bambi:2011yz,Matsas:2007bj,Richartz:2008xm,Richartz:2011vf,Yang:2022yvq}. In literature\cite{Semiz:2015pna}, Semiz and others corrected the views on the counter-effect and superradiance. Their results suggest that black holes can capture a small number of particles, which could potentially disrupt the black hole's event horizon. In summary, when test particles or scalar fields scatter near the event horizon of black holes, due to quantum effects, extremal or near-extremal black holes might absorb some hazardous particles or scalar fields, causing the black hole to overspin. This is significant for exploring the internal structure of black holes.

Therefore, seeking a more physically realistic black hole to explore the potential disruption of its event horizon is a meaningful research endeavor. In the quest for astrophysical black holes existing in the universe, people believe that conventional black hole solutions cannot exist due to the cosmic environment not being purely a vacuum and steady state, but rather filled with matter and fields (like dark energy or dark matter). This necessitates further research into black hole solutions that align more closely with physical reality. Recently, a popular "mimicker" emerged, namely the hairy Kerr black hole solution\cite{Contreras1,Contreras2}. This rotating hairy Kerr black hole is constructed through gravitational decoupling, not using the Newman-Janis (NJ) algorithm to construct rotation, thus avoiding some drawbacks brought about by complex coordinate transformations inherent in the NJ algorithm\cite{Drake:1998gf}. Theoretical research has extensively studied the hairy Kerr black hole, as seen in references\cite{Yang:2022ifo,Wu:2023wld,Avalos:2023jeh}. Around the hairy Kerr black hole, additional matter sources create deviations from the Kerr black hole structure, aligning more with physical reality and of great interest to us. In the spacetime of the hairy Kerr black hole, the event horizon radius strongly depends on the hairy parameters( $\alpha,l_0$), and their laws satisfy the strong energy conditions outside the event horizon\cite{Ovalle:2020kpd}. We find that the presence of hairy parameters( $\alpha,l_0$) affects the inner and outer radius of the event horizon, sparking our intense interest in the internal structure of the hairy Kerr black hole. 
Because the hairy black hole spacetime is more in line with the physical reality, in this article, we will use test particles and scalar fields carrying large angular momentum to explore this spacetime.

The structure of the article is as follows: In the second section, we briefly introduce the hairy Kerr black hole. In the third and fourth sections, we use test particles and scalar fields, respectively, to examine the possibility of the event horizon disruption in the hairy Kerr black hole. In the final section, we provide a summary and further discussion. Throughout the article, we adopt the natural unit system where $c=G=1$.

\section{\label{sec:level2}Hairy black holes in gravitational decoupling}
In classical general relativity, a black hole can be described solely by its mass $M$, charge $Q$, and spin parameter $a$, as per the famous no-hair theorem\cite{Hawking:1971vc,Israel:1967wq}. However, like the weak cosmic censorship conjecture, this theorem cannot be described with a rigorous mathematical formula. To test its rigor, several articles have presented examples that contradict the no-hair theorem\cite{Herdeiro:2016tmi,Sotiriou:2013qea,Li:2015bfa,Herdeiro:2015tia}, with the first counterexample being the "charged" black hole solution found in literature\cite{Bizon:1990sr}. Additionally, in these counterexamples, people have discovered extra "hairs" such as scalar field hair and soft quantum hair\cite{Hawking:2016msc,Li:2015bfa,Herdeiro:2015tia}. In our real universe, black holes might be surrounded by dark matter or dark energy, or there could be additional hairs due to the interaction between black hole spacetime and matter, potentially leading to corresponding spacetime changes. Recently, Ovalle and others proposed a simple method for generating hairy black holes, which was later extended to rotating cases by Contreras et al\cite{Ovalle:2020kpd,Contreras1}. Through the method of gravitational decoupling, it is quite straightforward to extend many black holes from known solutions to non-trivial extensions.

For the extension of the Kerr black hole to hairy black holes, in Boyer-Lindquist coordinates, as given by\cite{Gurses:1975vu}
\begin{equation}
\begin{split}
ds^2=&\left[1-\frac{2r\widetilde{m}\left(r\right)}{{\widetilde{\rho}}^2}\right]dt^2+\frac{4\widetilde{a}r\widetilde{m}\left(r\right){sin}^2{\theta}}{{\widetilde{\rho}}^2}dtd\phi-\frac{{\widetilde{\rho}}^2}{\widetilde{\mathrm{\Delta}}}dr^2\\
&-{\widetilde{\rho}}^2d\theta^2-\frac{\widetilde{\mathrm{\Sigma}}{sin}^2{\theta}}{{\widetilde{\rho}}^2}d\phi^2,
\end{split}\label{1}
\end{equation}
where
\begin{equation}
{\widetilde{\rho}}^2=r^2+{\widetilde{a}}^2{cos}^2{\theta},
\label{2}
\end{equation}

\begin{equation}
\widetilde{\rho}^2=r^2+{\widetilde{a}}^2 cos^2{\theta},
\label{3}
\end{equation}

\begin{equation}
\widetilde{\mathrm{\Sigma}}=\left(r^2+{\widetilde{a}}^2\right)^2-{\widetilde{a}}^2\widetilde{\mathrm{\Delta}}{sin}^2{\theta},
\label{4}
\end{equation}

\begin{equation}
\widetilde{a}=\frac{\widetilde{J}}{\widetilde{M}}.
\label{5}
\end{equation}
In gravitational decoupling, the Einstein field equations can be written as
\begin{equation}
{\widetilde{G}}_{\mu\nu}=k{\widetilde{T}}_{\mu\nu}=k(T_{\mu\nu}+S_{\mu\nu}).
\label{6}
\end{equation}
There, $S_{\mu\nu}$ represents the additional source\cite{Ovalle:2017fgl}. Contreras et al. achieved a non-trivial extension of the Kerr black hole, namely the hairy Kerr black hole, by introducing an extra energy-momentum tensor into the known Einstein-Kerr black hole solution. The metric is as follows\cite{Contreras1}
\begin{equation}
\widetilde{m}=M-\alpha\frac{r}{2}e^{-r/(M-\frac{l_o}{2})},
\label{7}
\end{equation}

and
\begin{equation}
\widetilde{\Delta}=r^2+a^2-2rM+\alpha r^2e^{-r/(M-\frac{l_o}{2})}.
\label{8}
\end{equation}
The solution of this hairy Kerr black hole follows the strong energy condition and is similar to the Kerr black hole, with the singularity of spacetime existing at $g^{rr}=\widetilde{\Delta}=0$ and $\widetilde{\mathrm{\Sigma}}\neq0$. In the hairy Kerr black hole, $l_o=\alpha l$ represents the primary hair, with $\alpha$ being the deformation parameter caused by the presence of surrounding matter. The condition $l_o\le2M\equiv l_k$ ensures the asymptotic flatness of spacetime\cite{Contreras1}.

In our paper on hairy Kerr black holes, we primarily discuss the impact of the hair parameters ($\alpha,l_0$) on its event horizon. This is because we know that the event horizon of the hairy Kerr black hole is strongly dependent on the hair parameters ($\alpha,l_0$).

The event horizon of the hairy Kerr black hole is given by $g^{rr}=\widetilde{\Delta}=0$. That is, writing formula (\ref{8}) as 
\begin{equation}
\widetilde{\Delta}=r^2+a^2-2rM+\alpha r^2e^{-\frac{r}{M-\frac{l_o}{2}}}=0.
\label{9}
\end{equation}
From the calculation of Equation (\ref{9}), we obtain
\begin{equation}
r_h=M\pm M\sqrt{1-\frac{\gamma+a^2}{M^2}},
\label{10}
\end{equation}
there $\gamma = \alpha r_h^2 e^{-r_h/(M-\frac{l_o}{2})}$, and $r_h$ represents the event horizon. Analyzing equation (\ref{10}), it can be easily deduced that when $\gamma + a^2 < M^2$, the term under the square root is positive. From a physical perspective, this indicates a spacetime with a black hole. However, it is clear that when $\gamma+ a^2 > M^2$, the term under the square root becomes negative. In physical terms, this implies that the metric does not possess an event horizon, thus no longer describing a black hole. Our interest lies in the case where there is no event horizon, as this represents an opportunity to further understand the internal structure of black holes.

\begin{figure*}[ht]
    \centering
    \begin{subfigure}[b]{0.5\textwidth}
        \centering
        \includegraphics[width=.9\linewidth]{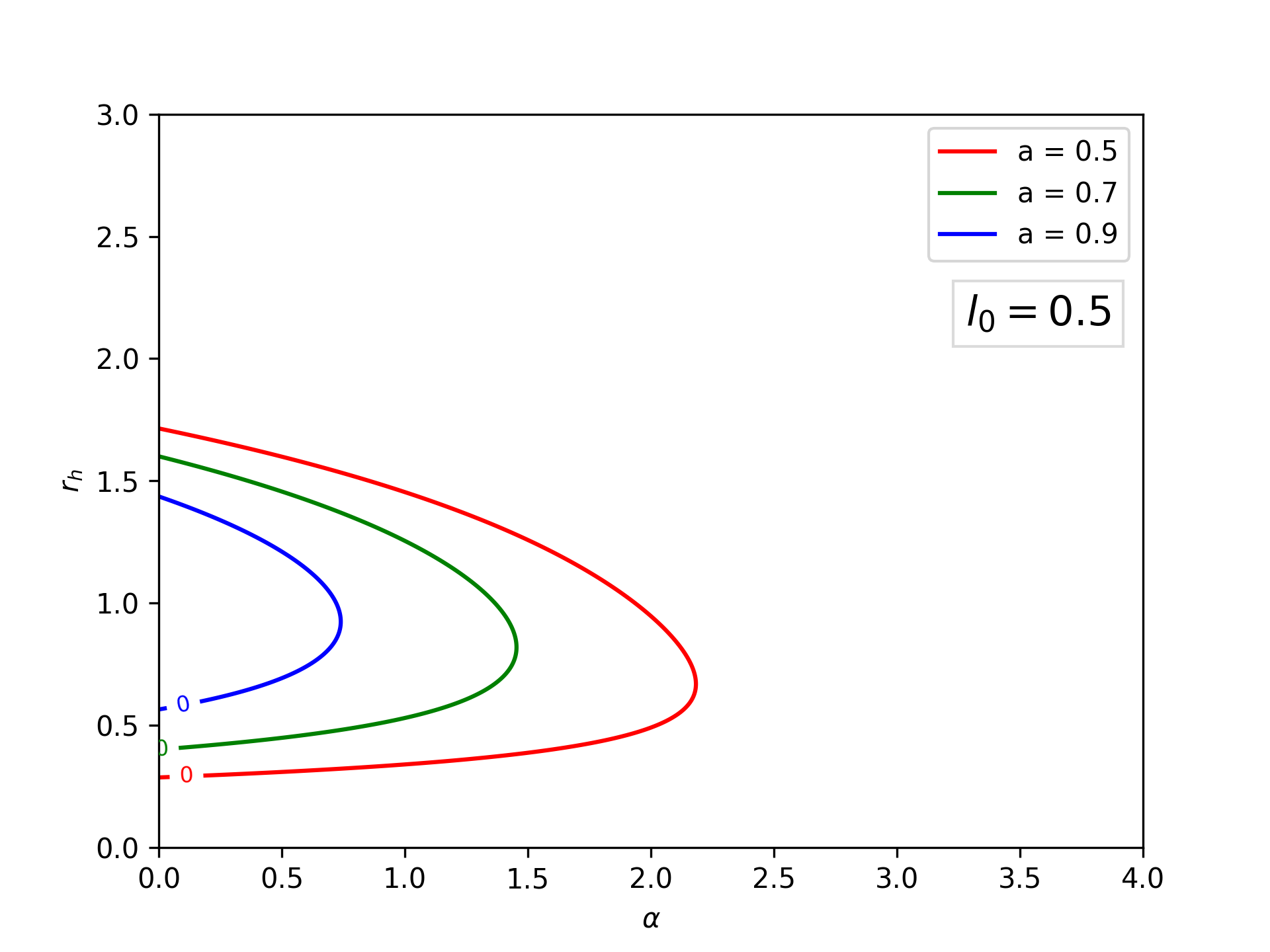}
        \caption{}
        \label{fig:sub1}
    \end{subfigure}%
    \begin{subfigure}[b]{0.5\textwidth}
        \centering
        \includegraphics[width=.9\linewidth]{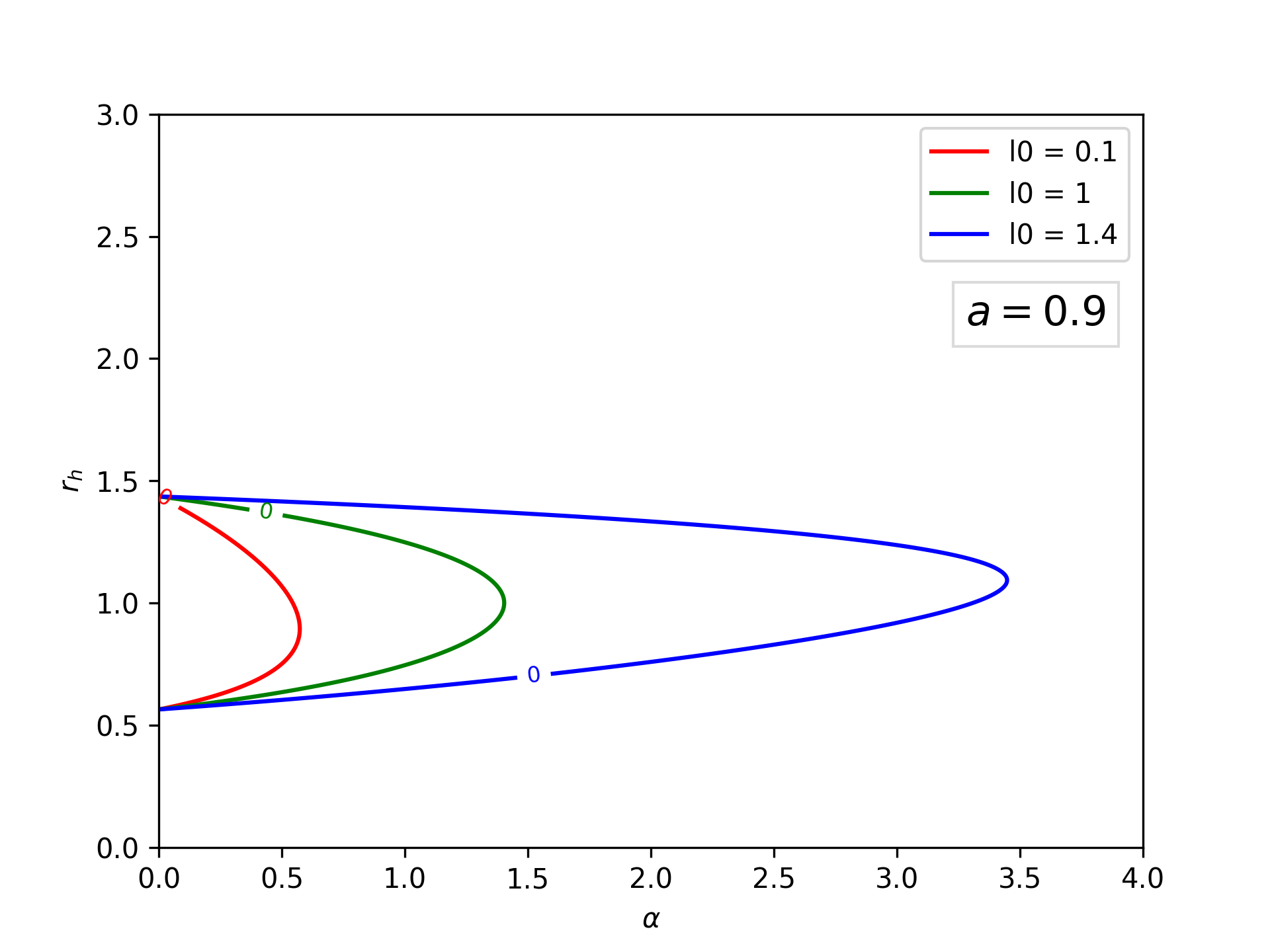}
        \caption{}
        \label{fig:sub2}
    \end{subfigure}
    \caption{(Figure \ref{fig:sub1}: The variation in the number of event horizons of the hairy Kerr black hole with respect to the hairy parameter $\alpha$ when the hairy parameter $l_0$ is set to a fixed value.
Figure \ref{fig:sub2}: The change in the number of event horizons of the hairy Kerr black hole with respect to the hairy parameter $\alpha$ when $l_0$ takes different values, with $M=1$.)}
    \label{fig:test}
\end{figure*}

To better describe the effect of the hairy parameters ($\alpha,l_o$) and the spin parameter $a$ on the hairy Kerr black hole, we will illustrate the relationship between the number of horizons of the hairy Kerr black hole and these parameters in a graph, labeled as Figure \ref{fig:test}. From the Figure \ref{fig:sub1} and Figure \ref{fig:sub2} in Figure \ref{fig:test}, we can intuitively discern that the number of event horizons is strongly dependent on the variation in the hairy parameters. For the Figure \ref{fig:sub1}, as the parameter $\alpha$ gradually increases, the number of event horizons transitions from two to one, and eventually, the horizon disappears. This indicates that the value of 
$\alpha$ leads to the overspin of the hairy Kerr black hole. In the Figure \ref{fig:sub2}, different values of $l_0$ result in various degrees of disruption to the hairy Kerr black hole. However, we find that when the value of $\alpha$ in the hairy parameter is as large as possible and 
$l_0$ is as small as possible, in this case, the event horizon of the hairy Kerr black hole is more easily destroyed. Overall, the destruction of the event horizon of the hairy Kerr black hole is strongly dependent on the hairy parameters.

For the hairy Kerr black hole, the area of the event horizon is given by the following formula:
\begin{equation}
A=\iint{\sqrt{g_{\theta\theta}g_{\varphi\varphi}}d\theta d\varphi}=4\pi\left(r_h^2+a^2\right),
\label{11}
\end{equation}
and the angular velocity at the event horizon is as follows:
\begin{equation}
\Omega_H=-\frac{g_{03}}{g_{33}}=\frac{a}{r_h^2+a^2}.
\label{12}
\end{equation}

\section{\label{sec:level3}Investigating the impact of hairy parameters on the destruction of the event horizon through test particles}
To investigate whether the hairy Kerr black hole can be overspun using test particles, we need to examine the following conditions: first, whether the test particles can fall into the event horizon; second, whether the test particles that fall into the event horizon meet the conditions for overspinning the hairy Kerr black hole. When both of these conditions are satisfied, we analyze the impact of hairy parameters on the destruction of the event horizon.

According to the calculations in Section \ref{sec:level2}, the equation for the event horizon of the hairy Kerr black hole is as follows:
\begin{equation}
r_h=M\pm M\sqrt{1-\frac{\gamma+a^2}{M^2}}
\label{13}
\end{equation}
In this equation, $\gamma=\alpha r_h^2e^{-r_h/(M-\frac{l_o}{2})}$, and $r_h$  represents the event horizon. Analyzing the equation (\ref{13}), the event horizon of the hairy Kerr black hole exists only when $\gamma+a^2\le M^2$. However, when $\gamma+a^2>M^2$, it implies that the event horizon does not exist in this spacetime, exposing the spacetime singularity at the center to observers at infinity. This provides an opportunity to understand the internal structure of black holes. In this paper, we primarily discuss this scenario.

In the spacetime of the hairy Kerr black hole, the motion of particles is typically described by the geodesic equation:
\begin{equation}
\frac{d^2x^\mu}{d\tau^2}+\Gamma_{\alpha\beta}^\mu\frac{dx^\alpha}{d\tau}\frac{dx^\beta}{d\tau}=0,
\label{14}
\end{equation}
The Lagrangian for this system is:
\begin{equation}
L=\frac{1}{2}\mu g_{\mu\nu}\frac{dx^\mu}{d\tau}\frac{dx^\nu}{d\tau}
=\frac{1}{2}\mu g_{\mu\nu}{\dot{x}}^\mu{\dot{x}}^\nu.
\label{15}
\end{equation}

To measure the angular momentum and energy of test particles, we design the test particles to slowly approach along the equatorial plane. Since the test particles move on the equatorial plane, there is no motion in the $\theta$ direction, which means $\frac{d\theta}{d\tau}=0$. Therefore, the momentum of the test particle in the $\theta$ direction is zero, and the test particle only moves on the equatorial plane, with no component in the $\theta$ direction (i.e., $\frac{d\theta}{d\tau}=0$), which can be expressed as
\begin{equation}
p_{\theta}=\frac{\partial L}{\partial\dot{\theta}}=mg_{22}\dot{\theta}=0.
\label{16}
\end{equation}
In this case, based on the motion equation of the test particle, the angular momentum $\delta J$ and energy $\delta E$ can be represented by the components of the test particle in the $\phi$ and $t$ directions, respectively, as follows
\begin{equation}
\delta J=P_\phi=\frac{\partial L}{\partial\phi}=mg_{3\nu}{\dot{x}}^\nu,
\label{17}
\end{equation}
and
\begin{equation}
\delta E=-P_t=-\frac{\partial L}{\partial\dot{t}}=-mg_{0\nu}{\dot{x}}^\nu.
\label{18}
\end{equation}

If a test particle can enter the interior of the event horizon, then the total energy and angular momentum of the hairy Kerr black hole will also undergo corresponding changes. This means that when the test particle enters the interior of the event horizon, a composite system consisting of the hairy Kerr black hole is formed. At this point, the changes in the energy and angular momentum of this composite system are expressed as follows 
\begin{equation}
M\longrightarrow M^,=M+\delta E,
\label{19}
\end{equation}
and
\begin{equation}
J\longrightarrow J^,=J+\delta J.
\label{20}
\end{equation}

When a test particle moves outside the event horizon of a the hairy Kerr black hole, its four-dimensional velocity is timelike, represented by the following equation:
\begin{equation}
U^\mu U_\mu=\frac{dx^\mu}{d\tau}\frac{dx_\mu}{d\tau}=g_{\mu\nu}\frac{dx^\mu}{d\tau}\frac{dx^\nu}{d\tau}=\frac{1}{\mu^2}g^{\mu\nu}P_\mu P_\nu=-1.
\label{21}
\end{equation}
After rearranging this equation, we obtain
\begin{equation}
g^{00}{\delta E}^2-2g^{03}\delta E\delta J+g^{11}P_r^2+g^{33}{\delta J}^2=-m^2.
\label{22}
\end{equation}
From this, we can calculate the energy increment gained by the hairy Kerr black hole from the test particle, given by 
\begin{equation}
\delta E=\frac{g^{03}}{g^{00}}\delta J\pm\frac{1}{g^{00}}\sqrt{\left[{(g^{03})}^2{\delta J}^2-g^{00}(g^{33}{\delta J}^2+g^{11}P_r^2+m^2\right]}.
\label{23}
\end{equation}
Regarding the equation (\ref{23}) mentioned above, we find that the energy can have two values. However, careful analysis reveals that when considering a test particle moving from infinity towards the event horizon, the motion of the test particle must be timelike and future-directed. This requires satisfying the following condition: 
\begin{equation}
\frac{dt}{d\tau}>0.
\label{24}
\end{equation}
By organizing the energy and angular momentum of the test particle, we find that when the conditions of the motion being timelike and future-directed are met, the derived conditions for energy and angular momentum are as follows:
\begin{equation}
\delta E>-\frac{g_{03}}{g_{33}}\delta J.
\label{25}
\end{equation}
Therefore, under the energy condition of equation (\ref{25}), it can be concluded that the energy increment gained by the black hole from the test particle can only take the negative sign, expressed as 
\begin{equation}
\delta E=\frac{g^{03}}{g^{00}}\delta J-\frac{1}{g^{00}}\sqrt{\left[{(g^{03})}^2{\delta J}^2-g^{00}(g^{33}{\delta J}^2+g^{11}P_r^2+m^2\right]}
\label{26}
\end{equation}

In the analysis of the conditions that energy and angular momentum must satisfy, we find that for a test particle to precisely fall into the event horizon, its energy and angular momentum must fulfill the following condition
\begin{equation}
\delta J<-\lim_{r \to r_h} {\frac{g_{33}}{g_{03}}}\delta E.
\label{27}
\end{equation}
 By jointly solving equations (\ref{12}) and (\ref{27}), we obtain
 \begin{equation}
\delta J<-\lim_{r \to r_h} {\frac{g_{33}}{g_{03}}}\delta E=\frac{\delta E}{\Omega_H}=\frac{r^2+a^2}{a}\delta E.
\label{28}
\end{equation}
 This means that there must be an upper limit to the angular momentum of the test particle.From a physical intuition perspective, when a test particle has a very large angular momentum, the centrifugal repulsion becomes much greater than the attraction between the particle and the black hole. In this case, the test particle will 'deviate' from the hairy Kerr black hole. In other words, if the angular momentum of the test particle is too large, it will not be able to fall into the black hole. Therefore, for a test particle to be captured by the hairy Kerr black hole, there must be an upper limit to its angular momentum, denoted as $\delta J_{max}$. Hence, from equation (\ref{28}), we can determine that the upper limit of the angular momentum 
$\delta J$ of the test particle is
 \begin{equation}
\delta J_{max}<\frac{\delta E}{\Omega_H}=\frac{{r_h}^2+a^2}{a}\delta E.
\label{29}
\end{equation}

Furthermore, according to the event horizon conditions of the hairy Kerr black hole, we know that a test particle entering the event horizon does not imply that the hairy Kerr black hole can be overspun. This is because when a test particle enters the event horizon, it forms a composite system with the angular momentum and mass of the hairy Kerr black hole itself. The event horizon of the hairy Kerr black hole disappears only when the formed composite system satisfies $\gamma+a^2>M^2$. By rearranging this condition, we can derive 
 \begin{equation}
a>M\sqrt{1-\frac{\gamma}{M^2}}=M\sigma,
\label{30}
\end{equation}
that is 
 \begin{equation}
J>M^2\sigma.
\label{31}
\end{equation} 

When a test particle enters the event horizon, the hairy Kerr black hole absorbs the angular momentum $\delta J$ and energy $\delta E$ of the test particle, forming a new composite system. The condition for this system to disrupt the event horizon of the hairy Kerr black hole becomes 
\begin{equation}
J^,>{{\sigma^,M}^,}^2.
\label{32}
\end{equation} 
Here,$\sigma^,=\sigma-\frac{1}{2\sigma}\frac{\gamma\left(kM-2\right)}{M^3}\delta E-(O){\delta E}^2$ and there, $k=\frac{r_h}{{(2M-l_0)}^2}$. This is because $\sigma$ contains mass $M$, and when the composite system is formed, the mass in 
$\sigma$ also undergoes a corresponding change ($M+\delta E,\delta E\ll M$)(
A Taylor series expansion has been utilized here). Ignoring higher order small quantities, $\sigma'$ becomes $\sigma-\frac{1}{2\sigma}\frac{\gamma\left(kM-2\right)}{M^3}\delta E$, that is $\sigma^,=\sigma-\frac{1}{2\sigma}\frac{\gamma\left(kM-2\right)}{M^3}\delta E$, then equation (\ref{32}) becomes 
\begin{equation}
J^,>{{\sigma^,M}^,}^2={(\sigma-\frac{1}{2\sigma}\frac{\gamma\left(kM-2\right)}{M^3}\delta E)M}^{,2}.
\label{33}
\end{equation} 
By substituting equations (\ref{19}) and (\ref{20}) into equation (\ref{33}), we obtain 
\begin{equation}
J+\delta J>(\sigma-\frac{1}{2\sigma}\frac{\gamma\left(kM-2\right)}{M^3}\delta E){(M+\delta E)}^2.
\label{34}
\end{equation} 
 Expanding equation (\ref{34}) yields
\begin{equation}
J+\delta J>(\sigma-\frac{1}{2\sigma}\frac{\gamma\left(kM-2\right)}{M^3}\delta E)\left(M^2+{\delta E}^2+2M\delta E\right).
\label{35}
\end{equation} 
After rearranging, we get 
\begin{equation}
\begin{split}
\delta J>&\left({\sigma M}^2-J\right)+\left(2\sigma M-\frac{1}{2\sigma}\frac{\gamma\left(kM-2\right)}{M}\right)\delta E\\
&+(\sigma-\frac{1}{\sigma}\frac{\gamma\left(kM-2\right)}{M^2}){\delta E}^2-\frac{1}{2\sigma}\frac{\gamma\left(kM-2\right)}{M^3}{\delta E}^3.
\label{36}
\end{split}
\end{equation}
The equation (\ref{36}) represents a lower limit for a test particle to disrupt the event horizon of a hairy Kerr black hole. Analyzing the above equation, since $\delta E$ is a first-order small quantity, neglecting the effects of higher-order perturbations, equation (\ref{36}) becomes
\begin{equation}
\delta J_{min}>\left({\sigma M}^2-J\right)+\left(2\sigma M+\frac{\gamma}{\sigma M}-\frac{\gamma k}{2\sigma}\right)\delta E.
\label{37}
\end{equation} 
This equation represents the lower limit of angular momentum for a test particle to disrupt the event horizon after entering it. Through the above analysis, only when the chosen test particle satisfies both conditions of equations (\ref{29}) and (\ref{37}), can the event horizon of the hairy Kerr black hole potentially be disrupted, exposing the internal structure of the hairy Kerr black hole.

Next, we will discuss two scenarios separately: extremal and near-extremal cases. In the extremal case, when $\frac{\gamma+a^2}{M^2}=1$, the event horizon of a hairy Kerr black hole can be written as
\begin{equation}
r_h=M.
\label{38}
\end{equation} 
In the first-order approximation, the condition for the destruction of the event horizon of such a hairy black hole can be expressed as follows 
\begin{equation}
\delta J_{max}<\frac{\delta E}{\Omega_H}=\frac{{r_h}^2+a^2}{a}\delta E,
\label{39}
\end{equation}
and
\begin{equation}
\delta J_{min}>\left(2\sigma M+\frac{\gamma}{\sigma M}-\frac{\gamma k}{2\sigma}\right)\delta E.
\label{40}
\end{equation}
By combining equations (\ref{30}), (\ref{38}), and (\ref{39}), we can calculate 
\begin{equation}
\begin{split}
\delta J_{max}<&\frac{{r_h}^2+a^2}{a}\delta E\\
&=\frac{M^2\sigma^2+M^2\sigma^2+\gamma}{M\sigma}\delta E=\left(2M\sigma+\frac{\gamma}{M\sigma}\right)\delta E.
\end{split}
\label{41}
\end{equation}
From Equations (\ref{40}) and (\ref{41}), it can be visually observed that the angular momentum and energy of the test particle can simultaneously satisfy these two conditions. In other words, as long as the angular momentum of the test particle falls within the energy range constrained by these two conditions, the event horizon of the hairy Kerr black hole can be destroyed. When the hair parameters ($\alpha, l_0$) approach zero (i.e., $\alpha=0$), the hairy Kerr black hole degenerates into a Kerr black hole. The analysis results from the above equation are consistent with previous studies on Kerr black holes, which conclude that the event horizon of an extremal Kerr black hole cannot be destroyed by test particles\cite{Zhao:2023vxq}.

If we take into account second-order or higher-order small quantities, it does not affect the analysis results. Combining Equations (\ref{36}), (\ref{40}), and (\ref{41}), we have 
\begin{equation}
\begin{split}
\delta J_{max}-\delta J_{min}=&\frac{\gamma k}{2\sigma}\delta E-\left(\sigma-\frac{1}{\sigma}\frac{\gamma\left(kM-2\right)}{M^2}\right){\delta E}^2\\
&+\frac{1}{2\sigma}\frac{\gamma\left(kM-2\right)}{M^3}{\delta E}^3.
\end{split}
\label{42}
\end{equation}
Clearly, for the given equation (\ref{42}), it is evident that when higher-order perturbations are taken into account, the upper limit of the angular momentum for the test particle is always greater than the lower limit. The symbol $\gamma$ represents a function of the hairy parameters $\gamma(\alpha, l_0)$. In other words, the equation (\ref{42}) can only have a possibility of being less than zero when there are no hairy parameters present ($\delta J_{max}-\delta J_{min}<0$). At this time, the hairy Kerr black holes has become a standard Kerr black hole. In other cases, it is greater than zero ($\delta J_{max}-\delta J_{min}>0$), corresponding to the scenario where hairy Kerr black holes can be overspun.

For another scenario, namely the near-extremal case ($a\approx \sigma M$), the conditions for test particles to enter the event horizon and disrupt the event horizon become
\begin{equation}
\delta J_{max}<\frac{{r_h}^2+a^2}{a}\delta E,
\label{43}
\end{equation}
and
\begin{equation}
\delta J_{min}>\left(2\sigma M+\frac{\gamma}{\sigma M}-\frac{\gamma k}{2\sigma}\right)\delta E+({\alpha M}^2-J).
\label{44}
\end{equation}
For the aforementioned case where $a\approx \sigma M$, we can describe the degree of proximity using a dimensionless small parameter $\epsilon$, given by 
\begin{equation}
\frac{\gamma+a^2}{M^2}=1-\epsilon^2.
\label{45}
\end{equation}
The parameter $\epsilon$ is a number that approaches zero, i.e., $\epsilon\ll 0$. When $\epsilon=0$, the equation becomes an extremal case. Based on equations (\ref{43}) and (\ref{44}), it can be concluded that in order to destroy the event horizon of the spacetime in the near-extremal case, the condition for destroying its event horizon becomes
\begin{equation}
\frac{1}{\Omega_H}-2\sigma M-\frac{\gamma}{\sigma M}+\frac{\gamma k}{2\sigma}>0.
\label{46}
\end{equation}
The term ($\sigma M^2-J$) neglected in the above equation is a second-order small quantity. That is to say, If the conditions stated in the equation are satisfied, the event horizon of a hairy Kerr black hole can be destroyed in the near-extremal case.

Due to $\epsilon \ll 1$, performing some series expansions yields
\begin{equation}
r_h=M(1+\epsilon),
\label{47}
\end{equation}
and
\begin{equation}
a=M\left(\sigma-\frac{\epsilon^2}{2\sigma}+O(\epsilon^4)\right).
\label{48}
\end{equation}
By combining equations (\ref{45}), (\ref{46}), (\ref{47}), and (\ref{48}), we obtain the calculation result as 
\begin{equation}
\begin{split}
\frac{1}{\Omega_H}-2\sigma M-&\frac{\gamma}{\sigma M}+\frac{\gamma k}{2\sigma}\\
&=\frac{a\gamma k+4M^2\sigma\epsilon+\left(2\sigma M^2+\frac{\gamma}{\sigma}\right)\epsilon^2-O(\epsilon^4)}{2a\sigma M}.
\label{49}
\end{split}
\end{equation}
The above analysis only considers the case of first-order perturbation. If we take higher-order perturbations into account and combine equations (\ref{36}) and (\ref{49}), we obtain the expression as 
\begin{equation}
\begin{split}
\delta J_{max}-&\delta J_{min}\\
&=\frac{a\gamma k+4M^2\sigma\epsilon+\left(2\sigma M^2+\frac{\gamma}{\sigma}\right)\epsilon^2-O\left(\epsilon^4\right)}{2a\sigma}\delta E\\
&-\left({\sigma M}^2-J\right)-\left(\sigma-\frac{1}{\sigma}\frac{\gamma\left(kM-2\right)}{M^2}\right){\delta E}^2\\
&+\frac{1}{2\sigma}\frac{\gamma\left(kM-2\right)}{M^3}{\delta E}^3.
\label{50}
\end{split}
\end{equation}
Among them $\left(\sigma M^2-J\right)=\frac{M^2\epsilon^2}{2\sigma}-O\left(\varepsilon^4\right)$.
Equation (\ref{50}) can be expressed as
\begin{equation}
\begin{split}
\delta J_{max}&-\delta J_{min}\\
&=\frac{a\Upsilon k+4M^2\sigma\epsilon+\left(2\sigma M^2+\frac{\gamma}{\sigma}\right)\epsilon^2-O\left(\epsilon^4\right)}{2a\sigma M}\delta E\\
&-\frac{M^2\epsilon^2}{2\sigma}-\left(\sigma-\frac{1}{\sigma}\frac{\gamma\left(kM-2\right)}{M^2}\right){\delta E}^2\\
&+\frac{1}{2\sigma}\frac{\gamma\left(kM-2\right)}{M^3}{\delta E}^3+O\left(\varepsilon^4\right).
\label{51}
\end{split}
\end{equation}
According to the equation above, it is clear that in the near-extremal case, even under higher-order perturbations, the event horizon of a hairy Kerr black hole can be disrupted by test particles due to the presence of hair parameters ($\alpha, l_0$),i.e.,$\delta J_{max}-\delta J_{min}>0$. Moreover, the larger the hair parameter, the easier the disruption, which suggests that the existence of hair parameters seems to facilitate the formation of naked singularities. These hair parameters ($\alpha, l_0$) are represented by $\gamma$, i.e., $\gamma(\alpha,l_0)$. Of course, if the hairy Kerr black hole degenerates into a Kerr black hole($\delta J_{max}-\delta J_{min}>0$), the conclusion obtained here is consistent with previous research, which states that the event horizon of a Kerr black hole in the near-extremal case can be disrupted\cite{Zhao:2023vxq}.

It is worth noting that, whether in extremal or near-extremal conditions, even when higher-order terms are taken into account (these higher-order terms include the spacetime background and the self-energy of the test body), we find that, in extremal conditions, it is approximately from Equation (\ref{42}) that
\begin{equation}
\delta J_{max}-\delta J_{min}=\frac{\gamma k}{2\sigma}\delta E+(higher-order terms).
\label{a}
\end{equation}
In near-extremal conditions, it is approximately from Equation (\ref{51}) that
\begin{equation}
\begin{split}
\delta J_{max}-\delta J_{min}=\frac{a\Upsilon k}{2a\sigma M}\delta E
+(higher-order terms).
\label{b}
\end{split}
\end{equation}
From the above two equations, it can be intuitively obtained that our analysis results are dominated by a first-order term. When the spacetime background and the backreaction of the test body are sufficiently considered, these backgrounds will affect the higher-order terms of our analysis framework. However, due to the actual theory behind spacetime and its complexity, we cannot accurately calculate the specific expressions of these higher-order terms. In this regard, existing articles have also discussed this point (see reference \cite{Li:2013sea,Hubeny:1998ga}), where they found that when the test body is uncharged, the impact of these effects is extremely small (this is because the electromagnetic self-field backreaction effect, which contributes most to the backreaction, can be ignored at this time). Since our test body is also uncharged, our backreaction will be even smaller, which means that it has a very small impact on our analysis results, at least not affecting the first-order results. Of course, since these effects are extremely important for testing the weak cosmic censorship conjecture, we will try to obtain a more accurate range through numerical evolution in future work.

\section{\label{sec:level4}Using a scalar field to overspin a hairy Kerr black hole}
Another method commonly used to test the Weak Cosmic Censorship conjecture is to scatter a black hole with a massive scalar field and investigate whether the resulting composite system can lead to the disruption of the black hole's event horizon. In this subsection, we also adopt Semiz et al.'s approach and use a scalar field carrying large angular momentum to scatter a hairy Kerr black hole\cite{Semiz:2005gs}. In order to study the possibility of event horizon disruption, we examine two scenarios: an extremal case and a near-extremal case, and analyze the impact of the hairy parameters on the disruption of the event horizon in these two situations.
\subsection{\label{sec:level4.1}Scattering of scalar fields with mass}
When a scalar field is incident on a hairy Kerr black hole, scattering occurs. Assuming that the mass of this scalar field $\psi$ is given by $m$, the motion equation of the scalar field can be described by the Klein-Gordon equation, which is as follows:
\begin{equation}
\begin{split}
\nabla_\mu\nabla^\nu-\mu^2\psi=0.
\label{52}
\end{split}
\end{equation}
According to the definitions of covariant and contravariant, the equation can be written as
\begin{equation}
\begin{split}
\frac{1}{\sqrt{-g}}\partial_\mu\left(\sqrt{-g}g^{\mu\nu}\partial_\nu\psi\right)-\mu^2\psi=0.
\label{53}
\end{split}
\end{equation}
According to the metric (\ref{1}), we can calculate its determinant as
\begin{equation}
\begin{split}
g=detg_{\mu\nu}=-{\widetilde{\rho}}^4\sin^2{\theta}.
\label{54}
\end{split}
\end{equation}
The contravariant tensor of metric (\ref{1}) is given by the following expression: 
\begin{equation}
\begin{split}
g^{\mu\nu}=\frac{\Delta^{\mu\nu}}{g}.
\label{55}
\end{split}
\end{equation}
Substituting metric (\ref{1}) into the above equation (\ref{53}) yields
\begin{equation}
\begin{split}
&\frac{\left(r^2+a^2\right)^2-a^2\widetilde{\Delta}\sin^2{\theta}}{\widetilde{\Delta}\widetilde{\rho}^2}\frac{\partial^2\psi}{\partial t^2}-\frac{4a \widetilde{m} r }{\widetilde{\Delta}\widetilde{\rho}^2}\frac{\partial^2\psi}{\partial t\partial\phi}\\
&-\frac{1}{\widetilde{\rho}^2}\frac{\partial}{\partial r}\left(\widetilde{\Delta}\frac{\partial\psi}{\partial r}\right)-\frac{1}{\widetilde{\rho}^2\sin^2\theta}\frac{\partial}{\partial\theta}\left(\sin\theta\frac{\partial\psi}{\partial\theta}\right)\\
&-\frac{\widetilde{\Delta}-a^2\sin^2\theta}{\widetilde{\Delta}\widetilde{\rho}^2\sin^2\theta}\frac{\partial^2\psi}{\partial\phi^2}-\mu^2\psi=0.
\label{56}
\end{split}
\end{equation}
The form of the solution for the scalar field $\psi$ in the above equation is as follows
\begin{equation}
\begin{split}
\psi\left(t,r,\theta,\phi\right)=e^{-i\omega t}R\left(r\right)S_{lm}(\theta)e^{im\phi}.
\label{57}
\end{split}
\end{equation}
The $S_{lm}(\theta)$ in the above equation is the angular spherical function, where $l$ and $m$ are constants for the angular separation variable, taking positive integer values. Substituting equation (\ref{57}) into the scalar field equation (\ref{56}) yields the angular equation for the scalar field:
\begin{equation}
\begin{split}
&\frac{1}{\sin^2{\theta}}\frac{d}{d\theta}\left(\sin{\theta}\frac{dS_{lm}\left(\theta\right)}{d\theta}\right)\\
&-\left(a^2\omega^2\sin^2{\theta}+\frac{m^2}{\sin^2{\theta}}-\mu^2a^2\cos^2{\theta}-\lambda_{lm}\right)S_{lm}\left(\theta\right)=0.
\label{58}
\end{split}
\end{equation}
The radial equation for the scalar field is obtained as
\begin{equation}
\begin{split}
&\frac{d}{dr}\left(\widetilde{\Delta}\frac{dR}{dr}\right)\\
&+\left(\frac{(r^2+a^2)^2}{\widetilde{\Delta}}\omega^2-\frac{4a \widetilde{m}r}{\widetilde{\Delta}}m\omega+\frac{m^2a^2}{\widetilde{\Delta}}+\mu^2r^2+\lambda_{lm}\right)R(r)=0.
\label{59}
\end{split}
\end{equation}
Upon solving equation (\ref{59}), it is found that its solution is a spherical function. Due to the normalization of spherical functions, when calculating the energy flux in the subsequent steps, integration over the entire event horizon surface is performed, with the integration of the spherical function equal to one. Therefore, we are now more concerned with the radial solution of the scalar field equation. For convenience in solving, we introduce the tortoise coordinate $r_\ast$ and define the tortoise coordinate as
\begin{equation}
\begin{split}
\frac{dr}{dr_\ast}=\frac{\widetilde{\Delta}}{r^2+a^2}.
\label{60}
\end{split}
\end{equation}
By substituting the tortoise coordinate from the above equation into the radial equation (\ref{60}) of the scalar field . we can resolve it to obtain
\begin{equation}
\begin{split}
&\frac{\widetilde{\Delta}}{\left(r^2+a^2\right)^2}\frac{d}{dr}\left(r^2\right)\frac{dR}{dr_\ast}+\frac{d^2R}{d{r_\ast}^2}+\left[\left(\omega-\frac{ma}{r^2+a^2}\right)^2+\right.\\
&\left. \frac{2\widetilde{\Delta}am\omega}{\left(r^2+a^2\right)^2}-\frac{\widetilde{\Delta}}{\left(r^2+a^2\right)^2}\left(\mu^2r^2+\lambda_{lm_0}\right)\right]R=0.
\label{61}
\end{split}
\end{equation}
We mainly analyze the vicinity of the event horizon ($r\sim r_h$), where
\begin{equation}
\widetilde{\Delta}\cong 0.
\label{62}
\end{equation}
By substituting equation (\ref{62}) into equation (\ref{61}), we can approximate it as follows
\begin{equation}
\begin{split}
\frac{d^2R}{d{r_\ast}^2}+\left(\omega-\frac{ma}{r^2+a^2}\right)^2R=0.
\label{63}
\end{split}
\end{equation}
The expression for the angular velocity at the event horizon of a Kerr black hole is given by
The expression for the angular velocity at the event horizon of a Kerr black hole is given by
\begin{equation}
\Omega_H=\frac{a}{{r_h}^2+a^2}.
\label{64}
\end{equation}
By substituting equation (\ref{64}) into equation (\ref{63}), the radial equation for the scalar field can be expressed as
\begin{equation}
\frac{d^2R}{d{r_\ast}^2}+\left(\omega-m\Omega_H\right)^2R=0.
\label{65}
\end{equation}
The solution to equation (\ref{65}) can be written in exponential form as
\begin{equation}
R(r)~exp\left[\pm i(\omega-m\Omega_H)r_\ast\right].
\label{66}
\end{equation}
The positive and negative signs in the solution of equation (\ref{66}) correspond to outgoing and incoming waves, respectively. We primarily consider the scattering of a scalar field onto a hairy Kerr black hole, where the black hole absorbs the energy of the scalar field. Therefore, selecting the negative sign in equation (\ref{66}) is more physically realistic. The solution to the radial equation for the scalar field in this case is
\begin{equation}
R\left(r\right)=exp\left[-i(\omega-m\Omega_H)r_\ast\right].
\label{67}
\end{equation}
Substituting equation (\ref{57}) into equation (\ref{67}), we can obtain the approximate solution of the scalar field near the event horizon as
\begin{equation}
\psi\left(t,r,\theta,\phi\right)=exp\left[-i(\omega-m\Omega_H)r_\ast\right]e^{-i\omega t}S_{lm}(\theta)e^{im\phi}.
\label{68}
\end{equation}
With this approximate solution, we can proceed to calculate the angular momentum and energy absorbed by the hairy Kerr black hole when a scalar field scatters onto it. The absorbed energy and angular momentum can be obtained by calculating the flux of energy and angular momentum around the event horizon.

The energy-momentum tensor of a scalar field ($\psi$) with mass ($\mu$) can be expressed in the following form
\begin{equation}
T_{\mu\nu}=\partial_\mu\psi\partial_\nu\psi^\ast-\frac{1}{2}g_{\mu\nu}\left(\partial_\mu\psi\partial^\nu\psi^\ast+\mu^2\psi\psi^\ast\right).
\label{69}
\end{equation}
By substituting the metric (\ref{1}) into the above equation, we can obtain
\begin{equation}
T_t^r=\frac{{r_h}^2+a^2}{{\widetilde{\rho}}^2}\omega\left(\omega-m\Omega_H\right)S_{lm}(\theta)e^{im\phi}{S^\ast}_{l^,m^,}(\theta)e^{-im\phi},
\label{70}
\end{equation}
and
\begin{equation}
T_\phi^r=\frac{{r_h}^2+a^2}{{\widetilde{\rho}}^2}m\left(\omega-m\Omega_H\right)S_{lm}(\theta)e^{im\phi}{S^\ast}_{l^,m^,}(\theta)e^{-im\phi}.
\label{71}
\end{equation}
Therefore, the energy flux through the event horizon is given by
\begin{equation}
\frac{dE}{dt}=\iint{T_t^r\sqrt{-g}}d\theta d\phi=\omega(\omega-m\Omega_H)\left[{r_h}^2+a^2\right],
\label{72}
\end{equation}
and the angular momentum flux through the event horizon is given by
\begin{equation}
\frac{dJ}{dt}=\iint{T_\phi^r\sqrt{-g}}d\theta d\phi=m\left(\omega-m\Omega_H\right)\left[{r_h}^2+a^2\right].
\label{73}
\end{equation}
From the above two equations, it can be observed that when ($\omega > m\Omega_H$), the values of the angular momentum flux and energy flux through the horizon are positive. This implies that in this scenario, the hairy Kerr black hole extracts energy ($\delta E$) and angular momentum ($\delta J$) from the scalar field. When ($\omega < m\Omega_H$), however, the angular momentum flux and energy flux through the event horizon are negative. In this case, the scalar field extracts energy from the black hole, which is known as black hole superradiance\cite{Brito:2015oca}.

For a very small time interval (dt), the amount of angular momentum and energy absorbed by the spacetime from the scalar field can be expressed as
\begin{equation}
dE=\omega(\omega-m\Omega_H)\left[{r_h}^2+a^2\right]dt,
\label{74}
\end{equation}
and
\begin{equation}
dJ=m(\omega-m\Omega_H)\left[{r_h}^2+a^2\right]dt.
\label{75}
\end{equation}
By utilizing equations (\ref{74}) and (\ref{75}), we can determine the energy and angular momentum extracted by the black hole from the scalar field after the scattering process. With this acquired energy and angular momentum, we can analyze the disruption of the event horizon of the hairy Kerr black hole following the scattering of the scalar field.
\subsection{\label{sec:level4.2}The overspinning state of the hairy Kerr black hole after scattering with a scalar field}
In this section, we mainly analyze whether a composite system formed by the scattering of a scalar field carrying large angular momentum with a hairy Kerr black hole can disrupt its event horizon. We also discuss the impact of the hairy parameter on the disruption of the event horizon.

For a continuous scalar field scattering process, we approach it using the concept of infinitesimal differentials, analyzing each time interval $dt$. In the analysis process, each $dt$ interval is treated in the same way, with the only difference being the initial state parameter values. Therefore, we only need to analyze one specific process among them.

From the analysis in Section \ref{sec:level2}, it is known that the hairy Kerr spacetime can be destroyed when the spin parameter $a$ changes. Using the same approach, in the scalar field scattering process, the initial mass and angular momentum of the hairy Kerr spacetime are $M$ and $J$, respectively. After the scalar field scatters, the mass of the hairy Kerr black hole becomes $M^,=M+dE$, and the angular momentum becomes $J^,=J+dJ$, forming a composite system. By analyzing the formula of the event horizon, we can determine whether the event horizon is disrupted by examining the sign of the composite system $ \sigma'M'^2-J'$. If the sign is positive, the event horizon of the black hole always exists. If the sign of 
$ \sigma'M'^2-J'$ is negative, the event horizon of this type of Kerr black hole is disrupted, exposing its internal structure to observers at infinity.

For the changes of a composite system within a very short time interval $dt$, the energy and angular momentum of the composite system formed after a hairy Kerr black hole absorbs energy $\delta E$ and angular momentum $\delta J$ from the scalar field within this time interval $dt$ are
\begin{equation}
\begin{split}
\sigma^,{M^,}^2-J^,&=\left(\sigma-\frac{1}{2\sigma}\frac{\gamma\left(kM-2\right)}{M^3}\delta E\right)\left(M+dE\right)^2\\
&-\left(J+dJ\right)\\
&=\left({\sigma M}^2-J\right)+{\sigma dE}^2+2\sigma MdE\\
&-\frac{1}{2\sigma}\frac{\gamma\left(kM-2\right)}{M}dE-\frac{1}{\sigma}\frac{\gamma\left(kM-2\right)}{M^2}{dE}^2\\
&-\frac{1}{2\sigma}\frac{\gamma\left(kM-2\right)}{M^3}{dE}^3-dJ.
\label{76}
\end{split}
\end{equation}
When considering only the low-order terms and neglecting the higher-order terms, the above equation becomes
\begin{equation}
\begin{split}
\sigma^,{M^,}^2-J^,=\left(\alpha M^2-J\right)+\left(2\sigma M+\frac{\gamma}{M\sigma}-\frac{\gamma k}{2\sigma}\right)dE-dJ.
\label{77}
\end{split}
\end{equation}
Substituting the expressions for the energy and angular momentum absorbed by a hairy Kerr spacetime from the scalar field (Equations (\ref{74}) and (\ref{75})) analyzed in Section \ref{sec:level4.1} into Equation (\ref{77}), we obtain
\begin{equation}
\begin{split}
\sigma^,{M^,}^2-J^,&=\left(\sigma M^2-J\right)+\left(2\sigma M+\frac{\gamma}{M\sigma}-\frac{\gamma k}{2\sigma}\right)m^2\\&\times\left(\frac{\omega}{m}-\frac{1}{\left(2\sigma M+\frac{\gamma}{M\sigma}-\frac{\gamma k}{2\sigma}\right)}\right)\left(\frac{\omega}{m}-\mathrm{\Omega}_H\right)\\
&\times\left[{r_h}^2+a^2\right]dt.
\label{78}
\end{split}
\end{equation}

In extremal situations, that is $\alpha M^2=J$, equation (\ref{78}) becomes
\begin{equation}
\begin{split}
\sigma^,{M^,}^2-J^,&=\left(2\sigma M+\frac{\gamma}{M\sigma}-\frac{\gamma k}{2M\sigma}\right)m^2\\
&\times\left(\frac{\omega}{m}-\frac{1}{\left(2\sigma M+\frac{\gamma}{M\sigma}-\frac{\gamma k}{2\sigma}\right)}\right)\left(\frac{\omega}{m}-\mathrm{\Omega}_H\right)\\
&\times\left[{r_h}^2+a^2\right]dt.
\label{79}
\end{split}
\end{equation}
By using the formula for angular velocity (Equation (\ref{12})), in the extremal case, the angular velocity at the event horizon of a hairy Kerr black hole can be simplified as
\begin{equation}
\begin{split}
\Omega_H&=\frac{a}{r_h^2+a^2}=\frac{M\sigma}{M^2\sigma^2+M^2\sigma^2+\gamma}\\
&=\frac{1}{2M\sigma+\frac{\gamma}{M\sigma}}\le\frac{1}{\left(2\sigma M+\frac{\gamma}{M\sigma}-\frac{\gamma k}{2\sigma}\right)}.
\label{80}
\end{split}
\end{equation}
The equality in the above equation holds only when the hairy parameters ($\alpha, l_0$) are zero ($\alpha = 0$). From the above equation, it can be seen that the presence of other hairy fields or matter around a hairy Kerr black hole can have an impact, leading to a deviation in the angular velocity of the hairy Kerr black hole. This result also implies the possibility of disruption in the event horizon of the hairy Kerr black hole.

If we choose the mode of the incident scalar field as follows
\begin{equation}
\begin{split}
\frac{\omega}{m}=\frac{1}{2}\left(\frac{1}{\left(2\sigma M+\frac{\gamma}{M\sigma}-\frac{\gamma k}{2\sigma}\right)}+\mathrm{\Omega}_H\right).
\label{81}
\end{split}
\end{equation}
In that case, the state of the composite system becomes
\begin{equation}
\begin{split}
\sigma^,{M^,}^2&-J^,=-\frac{1}{4}\left(2\sigma M+\frac{\gamma}{M\sigma}-\frac{\gamma k}{2\sigma}\right)m^2\\
&\times\left(\frac{1}{\left(2\sigma M+\frac{\gamma}{M\sigma}-\frac{\gamma k}{2\sigma}\right)}-\mathrm{\Omega}_H\right)^2\left({r_h}^2+a^2\right)dt^2.
\label{82}
\end{split}
\end{equation}
According to formula (\ref{82}), it is evident that whether a hairy Kerr black hole can be destroyed in near extremal cases strongly depends on the hairy parameters ($\alpha, l_0$). We can obtain the following equation as the hairy parameters ($\alpha, l_0$) change, that is
\begin{equation}
\begin{split}
\sigma^,{M^,}^2-J^,\le0.
\label{83}
\end{split}
\end{equation}
The equality in the above equation holds only when the hairy parameters ($\alpha, l_0$) vanish, indicating that there are no additional hairs present in the spacetime. This represents the degeneration of the hairy Kerr black hole into a standard Kerr black hole, and the analytical results align with those obtained from general relativity, which states that extremal Kerr black holes cannot be disrupted by scalar fields.

In fact, due to the influence of the hairy parameters ($\alpha, l_0$), the angular velocity of the hairy Kerr black hole undergoes a deviation. This indirectly results in the existence of a range of wave modes capable of disrupting the event horizon, the range of which is analyzed in Equation (\ref{79}) as
\begin{equation}
\begin{split}
\Omega_H=\frac{1}{2M\sigma+\frac{\gamma}{M\sigma}}<\frac{\omega}{m}<\frac{1}{\left(2\sigma M+\frac{\gamma}{M\sigma}-\frac{\gamma k}{2\sigma}\right)}.
\label{84}
\end{split}
\end{equation}
This also implies that as long as the range of the incident wave modes satisfies the condition in the above equation, the composite system formed after the scalar field incident the black hole will satisfy
\begin{equation}
\begin{split}
\sigma^,{M^,}^2-J^,<0.
\label{85}
\end{split}
\end{equation}
In other words, when the mode range of the scalar field satisfies Equation (\ref{84}), the event horizon of an extremal hairy Kerr black hole can be disrupted. Moreover, the larger the values of the hairy parameters ($\alpha, l_0$), the greater the range of wave modes that satisfy this condition. This implies that it is easier to disrupt the event horizon of a hairy Kerr black hole as the values of the hairy parameters increase.

 In near extremal situations, i.e. $\sigma M^2\sim J$. In this case, there is the following equation
\begin{equation}
\begin{split}
\sigma^,{M^,}^2-J^,&=\left(\sigma M^2-J\right)+\left(2\sigma M+\frac{\gamma}{M\sigma}-\frac{\gamma k}{2M\sigma}\right)m^2\\
&\times\left(\frac{\omega}{m}-\frac{1}{\left(2\sigma M+\frac{\gamma}{M\sigma}-\frac{\gamma k}{2\sigma}\right)}\right)\left(\frac{\omega}{m}-\mathrm{\Omega}_H\right)\\
&\times\left[{r_h}^2+a^2\right]dt.
\label{86}
\end{split}
\end{equation}
If the mode of the incident scalar field is
\begin{equation}
\begin{split}
\frac{\omega}{m}=\frac{1}{2}\left(\frac{1}{\left(2\sigma M+\frac{\gamma}{M\sigma}-\frac{\gamma k}{2\sigma}\right)}+\mathrm{\Omega}_H\right).
\label{87}
\end{split}
\end{equation}
Then Equation (\ref{86}) becomes
\begin{equation}
\begin{split}
\sigma^,{M^,}^2-J^,&=\left(\sigma M^2-J\right)-\frac{1}{4}\frac{\mathrm{\Omega}_H}{\left(2\sigma M+\frac{\gamma}{M\sigma}-\frac{\gamma k}{2\sigma}\right)}m^2\\
&\times\left(\frac{1}{\mathrm{\Omega}_H}-\left(2\sigma M+\frac{\gamma}{M\sigma}-\frac{\gamma k}{2\sigma}\right)\right)^2\\
&\times\left({r_h}^2+a^2\right)dt^2.
\label{88}
\end{split}
\end{equation}

Similar to the consideration of the near-extremal case in Section \ref{sec:level2}, here we also consider using a dimensionless small quantity to represent its degree of approximation, i.e
\begin{equation}
\begin{split}
\frac{a^2+\gamma}{M^2}=1-\epsilon^2.
\label{89}
\end{split}
\end{equation}
Since $\epsilon$ is a number approaching zero, by expanding equation (\ref{89}) using the Taylor series and then substituting it into equation (\ref{88}), we have
\begin{equation}
\begin{split}
\left(\sigma M^2-J\right)=\frac{M^2\epsilon^2}{2\sigma}-O\left(\varepsilon^4\right),
\label{90}
\end{split}
\end{equation}
and
\begin{equation}
\begin{split}
\frac{1}{\mathrm{\Omega}_H}&-\left(2\sigma M+\frac{\gamma}{M\sigma}-\frac{\gamma k}{2\sigma}\right)\\
&=\frac{2\sigma\left[2M^2\epsilon+{M^2\epsilon}^2-O\left(\epsilon^4\right)\right]+a\gamma k+\frac{\gamma}{\sigma}\epsilon^2-O(\epsilon^4)}{2a\sigma}\\
&=\frac{a\Upsilon k+4M^2\sigma\epsilon+\left(2\sigma M^2+\frac{\gamma}{\sigma}\right)\epsilon^2-O(\epsilon^4)}{2a\sigma}.
\label{91}
\end{split}
\end{equation}
Then, we can obtain the final expression as
\begin{equation}
\begin{split}
\sigma^,{M^,}^2&-J^,=\left(\sigma M^2-J\right)-\frac{1}{4}\frac{\mathrm{\Omega}_H}{\left(2\sigma M+\frac{\gamma}{M\sigma}-\frac{\gamma k}{2\sigma}\right)}{\mathrm{\Omega}_H}^2\\
&\times m^2\left(\frac{1}{\mathrm{\Omega}_H}-\left(2\sigma M+\frac{\gamma}{M\sigma}-\frac{\gamma k}{2\sigma}\right)\right)^2\\
&\times\left({r_h}^2+a^2\right)dt\\
&=\left[\frac{M^2\epsilon^2}{2\sigma}-O\left(\varepsilon^4\right)\right]-\frac{1}{4}\frac{\mathrm{\Omega}_H}{\left(2\sigma M+\frac{\gamma}{M\sigma}-\frac{\gamma k}{2\sigma}\right)}m^2\\
&\times\left(\frac{a\gamma k+4M^2\sigma\epsilon+\left(2\sigma M^2+\frac{\gamma}{\sigma}\right)\epsilon^2-O(\epsilon^4)}{2a\sigma}\right)^2\ \\
&\times\left({r_h}^2+a^2\right)dt.
\label{92}
\end{split}
\end{equation}

During the analysis of the scalar field impingement process, we consider the composite state within an extremely short time interval $dt$, which means that here $dt$ is a first-order small quantity. From an analysis of the above equation, it is understood that only when the hairy parameters ($\alpha,l_0$) are zero, we have ${{\alpha M}^,}^2-J^,>0$. Under these circumstances, the spacetime becomes a Kerr black hole, and the results of the analysis are consistent with the idea that a near-extremal Kerr black hole cannot be destroyed by a scalar field. Only when the hairy parameters are non-zero, an analysis of the above formula makes it easy to deduce that the first term is a second-order small quantity ($O(+\epsilon^2)$) and the second term is a first-order small quantity ($O(-dt)$). The final effect of this expression is a negative value,that is
\begin{equation}
\begin{split}
\sigma^,{M^,}^2-J^,<0.
\label{93}
\end{split}
\end{equation}
This indicates that, due to the presence of 'hair' around the hairy Kerr black hole, these hairs cause a shift in the angular velocity of the hairy Kerr black hole, providing a possibility for the disruption of its event horizon.

In summary, we discuss the disruption of the event horizon by the incidence of scalar fields carrying large angular momentum on extremal and near-extremal hairy Kerr black holes. We find that in both extremal and near-extremal situations, the event horizon of the hairy Kerr black hole can be disrupted by the scalar field, and the probability of this disruption strongly depends on the 'hair' surrounding it. The more 'hair' present around a hairy Kerr black hole, the easier it is for its event horizon to be disrupted.
\section{Discussion and conclusions}
In this article, we have thoroughly explored the anomalies of the Weak Cosmic Censorship Conjecture in hairy Kerr black holes. Specifically, when we carefully examine the destruction of the event horizon of hairy Kerr black holes using Gedanken experiments with test particles or scalar fields, we find that the destruction of the event horizon is strongly dependent on the hairy parameters. In our research, we conclude that it is possible to overspin hairy Kerr black holes using test particles, and that, whether in extremal or near-extremal conditions, hairy Kerr black holes seem to violate the Weak Cosmic Censorship Conjecture. When we explore hairy Kerr black holes with scalar fields carrying large angular momentum, our findings suggest that, in extremal or near-extremal conditions, the additional hair ($\alpha,\l_0$) present in hairy Kerr black holes leads to their overspinning.

We investigated the scenarios where test particles and scalar fields are incident on hairy Kerr black holes. Compared to standard Kerr black holes, we find that the event horizons of standard Kerr black holes without hair are not destroyed, neither in the extremal case with test particle detection nor in both extremal and near-extremal cases with scalar field detection. However, the event horizons of hairy Kerr black holes can be disrupted. This clearly indicates that the ability of hairy Kerr black holes to overspin strongly depends on the hairy parameters($\alpha,\l_0$). Additionally, for a regular black hole, since there is no singularity at its center, its overspinning is not protected by the Weak Cosmic Censorship Conjecture, which is a point of interest for us. This is because if hairy Kerr black holes can act as "simulators" of astrophysical black holes, then in astrophysics, we can allow accretion disks to autonomously absorb appropriate particles to overspin them, providing a natural laboratory for exploring quantum gravity inside black holes.

Certainly, the applicability of the Weak Cosmic Censorship Conjecture in hairy Kerr black holes can be verified by constraining the hairy parameters  ($\alpha,\l_0$). If we can effectively limit or constrain the values of these hairy parameters, it would help ensure the Weak Cosmic Censorship Conjecture is upheld in the context of hairy Kerr black holes. This discovery is not only crucial for our future research directions but may also reveal a profound connection between the black hole no-hair theorem and the Weak Cosmic Censorship Conjecture. Specifically, this implies that a deep understanding and constraint of the hairy parameters are essential not only for the validity of the Weak Cosmic Censorship Conjecture but might also offer new insights into the black hole no-hair theorem, thereby enhancing our understanding of the interplay and influence between these two fundamental theories.

\section{acknowledgements}
We acknowledge the anonymous referee for a constructive report that has significantly improved this paper. This work was supported by the  Special Natural Science Fund of Guizhou University (Grants
No. X2020068, No. X2022133) and the National Natural Science Foundation of China (Grant No. 12365008).


\end{document}